\documentclass[a4paper]{article}

\usepackage{INTERSPEECH2022}
\usepackage{amsmath}
\usepackage{amssymb}
\usepackage{enumitem}
\usepackage{bm}
\usepackage{xcolor}
\usepackage{multirow}

\newcommand{\argmin}[1]{\underset{#1}{\operatorname{arg}\,\operatorname{min}}\;}

\title{EPIC TTS Models: Empirical Pruning Investigations Characterizing Text-To-Speech Models}
\name{Perry Lam$^1$, Huayun Zhang$^2$, Nancy F. Chen$^2$, Berrak Sisman$^1$}
\address{
  $^{1}$Singapore University of Technology and Design \\
  $^{2}$Institute for Infocomm Research, A*STAR, Singapore
}
\email{perry\_lam@mymail.sutd.edu.sg, \{zhang\_huayun, nfychen\}@i2r.a-star.edu.sg, berraksisman@u.nus.edu
}

\begin{document}

\maketitle

\begin{abstract}

    Neural models are known to be over-parameterized, and recent work has shown that sparse text-to-speech (TTS) models can outperform dense models. Although a plethora of sparse methods has been proposed for other domains, such methods have rarely been applied in TTS. In this work, we seek to answer the question: what are the characteristics of selected sparse techniques on the performance and model complexity? We compare a Tacotron2 baseline and the results of applying five techniques. We then evaluate the performance via the factors of naturalness, intelligibility and prosody, while reporting model size and training time. Complementary to prior research, we find that pruning before or during training can achieve similar performance to pruning after training and can be trained much faster, while removing entire neurons degrades performance much more than removing parameters. To our best knowledge, this is the first work that compares sparsity paradigms in text-to-speech synthesis.

\end{abstract}
\noindent\textbf{Index Terms}: speech synthesis, text-to-speech, vocoder, sparsity, network pruning

\section{Introduction}

Biological neural networks are known to be \textit{sparse}, whereby the number of inputs per neuron are typically much smaller the number of neurons per layer, on the order of 1/30 to 1/40 \cite{optimal-synaptic}. In fact, such sparsity seems to be \textit{required} for pattern classifier learning in the cerebellum \cite{sparse-required}. However, in the world of artificial neural networks, models have developed in the opposite direction: they have grown in size and complexity, culminating in such architectures as the recent 530 billion parameter MT-NLG \cite{giant-model}. The sheer amount of data, hardware and time required for such models means that their utility has become limited, whereas their utility bill has become unlimited.

In response, researchers have tried to train sparse models which perform as well as dense models, showing that sparsity can not only improve performance, but also reduce overfitting \cite{overfitting} and increase robustness \cite{robustness}. To explain why this happens, the lottery ticket hypothesis has been proposed \cite{lottery-ticket}, which posits that a randomly initialized model contains subnetworks that are especially suited for training on the given task (i.e. \textit{winning tickets}), and that large models exponentially increase the chance of getting winning tickets.

To find these optimal subnetworks, various sparsification approaches have been proposed: (1) pruning pre-trained models, (2) training sparse models from scratch, and (3) reducing entire units altogether. The first approach includes Unstructured magnitude pruning (UMP) \cite{ump} or its derivatives Iterative Magnitude Pruning (IMP) \cite{lottery-ticket} and Prune-Adjust-Re-Prune (PARP). The second includes methods which prune before training like Single-shot Network Pruning (SNIP) \cite{snip} and Gradient Signal Preservation (GraSP) \cite{grasp}, or prune during training like Sparse Momentum (SM) \cite{sparse-momentum} and Rigged Lottery (RigL) \cite{rigged-lottery}. Finally, the third group subsume diverse methods such as structured trimming of neurons, using lower-rank tensor approximations by matrix factorization, and cross-layer parameter sharing.

These efforts have been applied and benchmarked on such fields as computer vision (CV) \cite{pruning-state}, natural language processing (NLP) \cite{pruning-nlp}, and speech recognition \cite{svd}, but surprisingly little has been done on TTS. Furthermore, the impressive sparsities attained in CV/NLP usually involve classification and not generation tasks \cite{pruning-state}, where the limited number of possible outputs and non-recurrent inputs may make many parameters redundant.

Thus, our main contribution is to show how generic sparsity-inducing methods can be applied in TTS, and characterize their associated tradeoffs in speech quality (in terms of intelligibility, naturalness and prosody) and model complexity (in terms of model size and training time).

The rest of this paper is organized as follows: Section 2 summarizes common sparse methods, Section 3 shows the experimental setup, Section 4 discusses our results and Section 5 concludes our paper.

\section{Related Work}

TTS is a seq2seq task that takes character or phoneme sequences as inputs and aims to generate audio waveforms. As waveforms are somewhat repeating but long and intractable, common practice is to transform the waveforms into mel spectrograms and use these spectrograms as target outputs for the TTS system. The waveforms can be reconstructed later via a separate vocoder.

As TTS systems improve, their parameter count has also grown -- for example, the original Tacotron \cite{tacotron} has roughly 7M parameters, while the updated Tacotron2 \cite{tacotron2} has about 28M. Extra features add more parameters, such as direct convolutional attention (DCA) \cite{dca}, which tweaks the attention mechanism to achieve the best tradeoff between length robustness, alignment speed and naturalness; and double decoder consistency \cite{ddc} (DDC), which includes an extra coarse decoder block to improve attention alignment.

To address this, the trend has been to propose lightweight models with lower memory requirements and faster inference suitable for mobile devices, but they either require designing new architectures like Device-TTS \cite{device-tts}, or extensive engineering to reduce existing models \cite{tacotron-lpc} \cite{taco-lpc}. For research models to be practical outside of the cloud, easily-applicable sparsity techniques are needed. Fortunately, unlike current NLP models with parameters on the order of $10^8$ and above, TTS models are at $10^7$ parameters or fewer, making them tractable and attractive to prune.

To our best knowledge, the only study on sparsifying TTS models so far is \cite{parp-tts}, where the authors applied their PARP technique, previously invented for speech recognition, on TTS. They discovered that both TTS systems (Transformer-TTS and Tacotron2) and vocoders (ParallelWaveGAN) are indeed highly prunable while maintaining similar or better performance compared to dense models, up to a sparsity of about 80\%. Additionally, they found data augmentation to have little impact on the pruned model's results, while knowledge distillation worsens it; this suggests that simple approaches work best. Inspired by their work, we explore other alternatives to prune models, which is the focus of the next section.

\section{Sparse Methods on TTS}

\subsection{Pruning pre-trained models}

The simplest approach for pruning is to start with a dense model and search for a binary mask $m$ that minimizes loss over the model's masked weights $w$, i.e. $\argmin{m} L(m \odot w)$, based on some criterion. Unstructured magnitude pruning (UMP) \cite{ump} ranks weights by their magnitude and retains the top-$k$ (where $k$ is the number of weights to keep). UMP can be combined with further training to achieve Iterative Magnitude Pruning (IMP) \cite{lottery-ticket}, where models are repeatedly pruned and trained with pruned weights receiving no updates. An extension of IMP, Prune-Adjust-Re-Prune (PARP) \cite{parp-tts} allows pruned weights to receive updates and has shown superior performance over both unpruned models and IMP in speech tasks. Variants of the above algorithms allow progressive pruning (where pruning starts at lower sparsities and increases to the target level at the final pruning step), which are useful at high sparsities. 

\subsection{Sparse training from scratch}

Nonetheless, pruning trained dense models has a key drawback. If the pretrained model is not available, we incur extra computation cost by the additional steps of pruning and fine-tuning after training the full model. This has led to techniques where sparse models can be trained from scratch. Taking cue from the lottery ticket hypothesis \cite{lottery-ticket}, \textit{foresight pruning} methods try to discover a winning ticket before training begins, while \textit{dynamic pruning} actively redistributes connections during training. 

\begin{itemize}[itemsep=0pt,leftmargin=*]
    \item \textbf{Single-shot Network Pruning (SNIP)} \cite{snip}, a foundational method in foresight pruning, identifies salient connections before the start of training and removes the least important ones. The saliency criterion is based on computing gradients with respect to the mask $m$ over (a mini-batch of) the dataset $\mathcal{D}$. Approximating with a real-valued $\bm{m}$, the effect $g_j$ of removing connection $j$ (i.e. setting $m_j$ from 1 to 0) on the loss is
    $$\Delta L_j(\bm{\mathrm{w}}); \mathcal{D} \approx g_j(\bm{\mathrm{w}}; \mathcal{D})
    = \left. \frac{\partial L(\bm{m} \odot \bm{\mathrm{w}}; \mathcal{D})}{\partial m_j} \right| _{\bm{m=1}}$$
    Only the top-k connections by magnitude are kept. Training then proceeds as usual, with the mask applied throughout.
    
    \item \textbf{Gradient Signal Preservation (GraSP)} \cite{grasp} extends SNIP by not only computing the importance of individual connections, but also estimating the dependencies between them by computing the gradient's Hessian $\bm{\mathrm{H}}$. Since weights are known to be coupled \cite{correlated-neurons}, this approach makes sense. The effect $\bm{\mathrm{S}}$ of perturbing initial weights $\bm{\mathrm{w}}_0$ by $\bm{\delta}$ is
    \begin{align*}
        \bm{\mathrm{S}}(\bm{\delta}) = \Delta L(\bm{w}_0 + \delta) - \Delta L(\bm{w}_0) \\
        = 2\delta^\top \nabla^2 L(\bm{w}_0) \nabla L(\bm{w}_0) + \mathcal{O}(\lVert \bm{\delta} \rVert ^2_2)
        &= 2\bm{\delta}^\top \bm{\mathrm{Hg}} + \mathcal{O}(\lVert \bm{\delta} \rVert ^2_2)
    \end{align*}
    The impact of removing weights $\bm{w}$ is $\bm{\mathrm{S}(-w)} \approx \bm{-w}\odot\bm{\mathrm{Hg}}$.
    
    \item \textbf{Sparse Momentum (SM)} \cite{sparse-momentum} was one of the first dynamic pruning algorithms that did not redistribute connections randomly. In SM, top-$k$ sparsity is first enforced by UMP. The exponentially smoothed momentum magnitude $\bm{\mathrm{M}}_i$ for each layer $i$ is tracked every epoch. At the end of each epoch, the smallest $p$\% of weights in each layer are pruned, and the missing weights in layer $i$ are regrown in proportion to $\bm{\mathrm{M}}_i$. The intuition behind this is that connections with high momentum (gradient) during training are more likely to be important and should be regrown; this is certainly more effective than randomly regrowing weights in evolutionary approaches, and also allows for layer importance. Similar to learning rate scheduling, $p$ is decayed as the mask converges.
    
    \item \textbf{Rigged Lottery (RigL)} \cite{rigged-lottery} saves on the computational cost of SM by defining the sparsities of each layer beforehand, then operating layer by layer instead of across all layers. Every $t$ iterations, the smallest $p$\% of active weights in each layer are pruned. Then, the inactive weights with highest gradients (instead of momentum) at the current step are regrown and initialized to zero to receive updates at the next iteration.
\end{itemize}

\subsection{Reducing model units}

A key disadvantage of the previous approaches is that current deep learning frameworks operate on dense tensors, i.e. model sparsity merely indicates the proportion of zeros in the tensors and does imply a reduced model size.
For a 2D float32 tensor, the overhead of storing nonzero index locations only makes sparse tensors save memory at sparsities over 80\%. Furthermore, sparse tensors are converted to dense tensors before matrix operations can be applied, which slows inference time.

Therefore, a separate line of research has focused on removing entire neurons \cite{structured-pruning} or convolutional channels \cite{channel-pruning} corresponding to tensor rows/columns, which is termed \textit{structured trimming}. Higher sparsities are harder to achieve vis-à-vis pruning, because salient connections are often scattered across neurons. \cite{structured-lottery} has evaluated trimming criteria for generative audio and concluded that the best criterion for removing neurons remains the total magnitude of input weights $\sum_{j=1}^{N_{in}}\lvert w_{i,j} \rvert$.

\begin{figure*}[t]
  \centering
  \includegraphics[width=\textwidth]{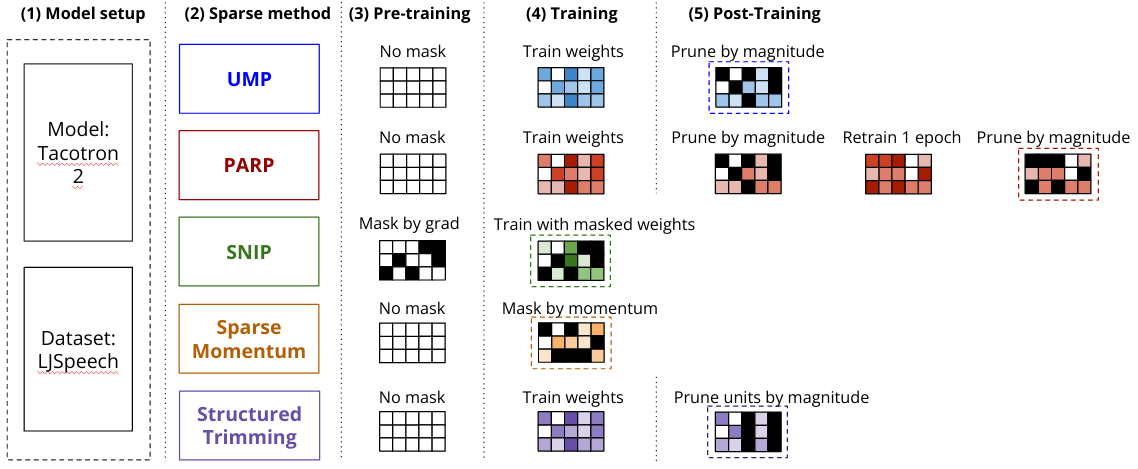}
  \setlength{\belowcaptionskip}{-10pt}
  \setlength{\abovecaptionskip}{-10pt}
  \caption{We compare the sparsity methods above, where the dotted boxes highlight the final pruning outputs.}
  \label{fig:experiments}
\end{figure*}

In the area of model compression, matrix factorization has been widely applied, often in speech recognition \cite{svd}. Using the singular value decomposition $\bm{\mathrm{A}}_{m\times n} = \bm{\mathrm{U}}_{m\times n} \bm{\mathrm{\Sigma}}_{n\times n} \bm{\mathrm{V}}_{n\times n}$, one can find a low-rank approximation to $\bm{\mathrm{A}}$ by zeroing the smallest elements in $\bm{\mathrm{\Sigma}}$. Alternatively, with the rise of Transformer-based models, parameter sharing across layers can enforce consistency between Transformer blocks and reduce total parameter count, e.g. in the language model ALBERT \cite{albert}. Finally, shapeshifter networks \cite{shapeshifter} allow cross-layer parameter sharing to be done automatically given a total parameter budget.


\section{Experiments}

In our work, we focus on pruning the text-to-melspectrogram network as it is the main determinant of speech quality. We select pruning techniques that represent each group of sparsity approaches: (1) UMP, (2) PARP (pruning pre-trained models), (3) SNIP (foresight pruning), (4) Sparse Momentum (dynamic pruning) and (5) Global Trimming (reducing model units).  These are summarized in Figure \ref{fig:experiments}.

\subsection{Baseline and Dataset}

For our baseline, we use the open-source Coqui-TTS\footnote{https://github.com/coqui-ai/TTS} framework to ensure standardization, maintainability, and reproducibility. We train Tacotron2 with DDC and DCA, and use ParallelWaveGAN \cite{parallelwavegan} for the vocoder.

Tacotron2 was trained for 100K steps on the default settings at batch size 32 on a modified stopnet loss to avoid repeating sequence generation. We use the LJSpeech dataset \cite{ljspeech} for all experiments and split it into train/dev/test sets at a ratio of 80:10:10. All our code is available on Github\footnote{https://github.com/iamanigeeit/coqui-tts}.

\subsection{UMP}
We apply UMP at multiple sparsities ranging from 10\% to 60\%, excluding the stopnet from pruning for its disproportionate effect on output termination.

\subsection{PARP}
Following \cite{parp-tts}, we applied UMP on our baseline at the target sparsities, then trained for 1 epoch (not to convergence!) and applied UMP again. From our experiments, we found that retaining the baseline optimizer helped to stabilize initial gradients post-pruning, and we set the scheduler at a constant learning rate of 5e-5, which we found to accelerate convergence. 

\subsection{SNIP}

We implemented SNIP by (1) randomly initializing Tacotron2, (2) saving the initial weights, (3) masking all learnable weights except for the stopnet and batch normalization layers, (4) computing the total gradients with respect to the mask across the training dataset, and (5) setting the mask to 0 for the lowest absolute gradients. Gradients are saved to produce masks of different sparsities. We accumulated gradients for the whole training dataset, instead of a mini-batch, to avoid sampling issues.

At the start of training, the model is loaded with the saved initial weights from (2). Training then proceeds with the same settings as our baseline, with the mask from (5) applied before every forward pass. However, because the mask identifies irrelevant connections early, training converges faster and the overall loss begins to increase before 100K steps; we therefore use the checkpoint with best dev loss for evaluation. 

\subsection{Sparse Momentum}
We used the \texttt{sparselearning} library\footnote{https://github.com/TimDettmers/sparse\_learning} from the original paper, but kept the scheduler and optimizer from our baselines. Our prune rate is set to 0.2, which stabilized gradients over the default 0.5. We experiment with sparsity levels of 20\% and 40\%. Similar to SNIP, we exclude the stopnet and batch normalization from the pruning process.


\subsection{Global Structured Trimming}
We followed \cite{structured-lottery} in removing units corresponding to tensor rows/columns with the lowest total magnitude. However, because of the heterogeneous nature of the Tacotron2 layers, comparing the importance of units across layers becomes dubious, even if layers are normalized by largest value or layer dimensionality as suggested in \cite{structured-lottery}. We therefore use the proportion of weights removed from each layer during UMP as a proxy for the proportion of units to be deleted per layer. 
As with previous settings, we do not trim the stopnet and batch normalizations, but here we also the ignore the linear projection layer and postnet final layer since they should output all 80 mel channels. 


\section{Results}
Following \cite{parp-tts}, we report our results on three measures of speech quality: naturalness, intelligibility and prosody. To compare the tradeoffs under each setting, we also report the model size and training time. All techniques were assessed at 20\% and 40\% sparsity, except for structured trimming, which were set to 2\% and 4\% due to its inherent difficulty.

\subsection{Intelligibility}

Intelligibility was measured by word error rate (WER) \cite{wer} when the generated audio was transcribed automatically from the Voicegain platform.

\begin{figure}[h]
  \centering
  \includegraphics[width=\linewidth]{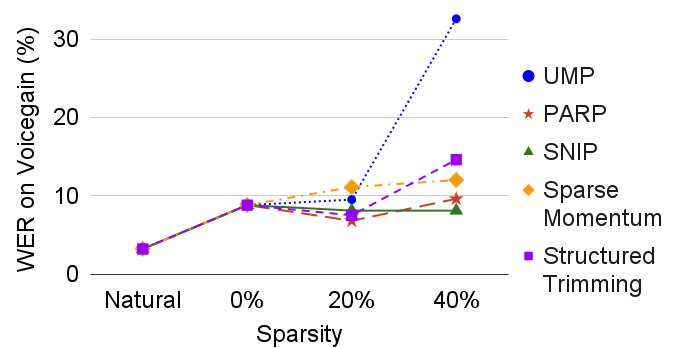}
  \setlength{\belowcaptionskip}{-5pt}
  \setlength{\abovecaptionskip}{-5pt}
  \caption{WER across settings weighted by ground truth length.}
  \label{fig:wer}
\end{figure}

In Figure \ref{fig:wer}, we see lower performance compared to \cite{parp-tts}, which we attribute to self-training the baseline instead of using a highly optimized pretrained model, using only LJSpeech in lieu of multiple datasets, and also performing PARP for only 1 epoch rather than to convergence. Note that it is possible to improve over the baseline with PARP and SNIP.

\subsection{Naturalness}
Naturalness was measured via mean opinion score (MOS) \cite{mos} from 1 (very unnatural) to 5 (very natural), on 10 randomly-chosen utterances across all techniques except UMP. Because of the closeness in WER for SNIP and Sparse Momentum with the baseline, we also did a three-way preference test for each of them between 0\% (baseline), 20\% and 40\% sparsities. Both tests were implemented on PsyToolkit \cite{psytoolkit-1} \cite{psytoolkit-2} and 15 full responses were obtained. We did not find significant differences between the masking methods, with Sparse Momentum slightly ahead; however, Structured Trimming was very challenging even at 2\% and 4\% (Table \ref{tab:mos}). While baselines were preferred against pruned models (Table \ref{tab:preference}), the difference was marginal, and participants noted that they often could not decide between samples.

\begin{table}[th]
    \caption{MOS. Bold indicates best method for given sparsity.}
    \label{tab:mos}
    \centering
  
    \begin{tabular}{r|c|c|c|c}
    \toprule
        \textbf{Sparse Method} & \textbf{Natural} & \textbf{0\%} & \textbf{20\%} & \textbf{40\%} \\
    \midrule
        PARP & & & 3.52 & 3.47 \\
        SNIP & & & \textbf{3.65} & 3.39 \\
        Sparse Momentum & & & \textbf{3.65} & \textbf{3.54} \\
        Structured Trimming & \multirow{-4}{*}{4.60} & \multirow{-4}{*}{3.64} & 3.34 & 3.13
    \end{tabular}
\end{table}
\vspace{-10pt}

\begin{table}[th]
    \caption{Preference test. Participants were given three audio samples at 0\%, 20\% and 40\% sparsity to indicate the most and least natural sample. Numbers below are percentage chosen.}
    \label{tab:preference}
    \centering
    \begin{tabular}{rccc}
        
        \toprule
        \textbf{Sparsity}      & \textbf{0\%}  & \textbf{20\%} & \textbf{40\%}\\
        \midrule
        SNIP (best)            & \textbf{38.7} & 36.7          & 24.7         \\
        SNIP (worst)           & \textbf{30.7} & 32.7          & 36.7         \\
        Sparse Momentum (best) & \textbf{39.3} & 32.7          & 28.0         \\
        Sparse Momentum (least)& \textbf{29.3} & 36.7          & 34.0        
    \end{tabular}
\end{table}

\vspace{-15pt}

  

\subsection{Prosody}
\begin{figure}[ht]
  \centering
  \includegraphics[width=\linewidth]{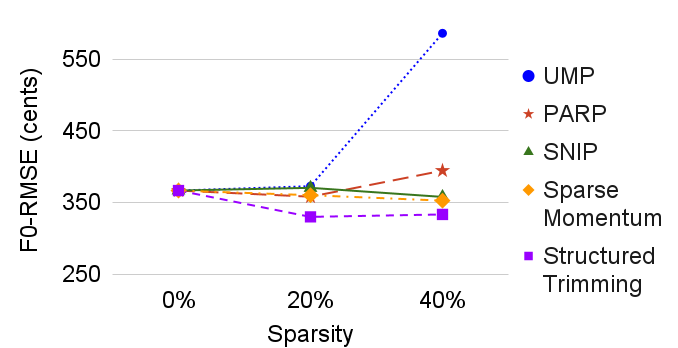}
  \setlength{\belowcaptionskip}{-10pt}
  \setlength{\abovecaptionskip}{-5pt}
  \caption{F0-RMSE (in cents) across all experiment settings.}
  \label{fig:f0}
\end{figure}

We used fundamental frequency root-mean-square error (F0-RMSE) against the ground truth audio as a proxy to quantify intonation differences according to the implementation of \cite{f0-rmse}. There were no significant differences between the baselines and sparse models, except for the UMP outlier (Figure \ref{fig:f0}).

\begin{figure}[h]
  \centering
  \includegraphics[width=\linewidth]{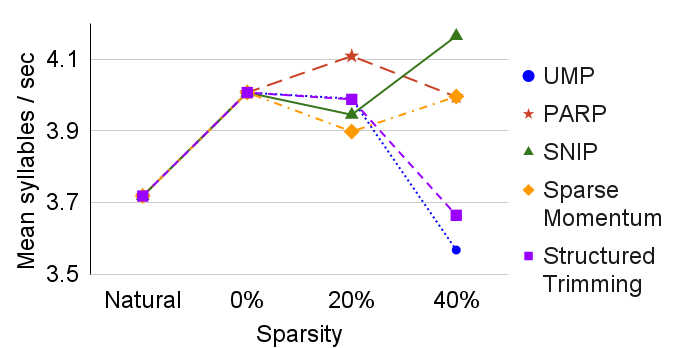}
  \setlength{\belowcaptionskip}{-5pt}
  \setlength{\abovecaptionskip}{-5pt}
  \caption{Syllable speed across all experiment settings.}
  \label{fig:syllable}
\end{figure}

To compare speaking rates, we compared the Voicegain transcript and the utterance duration over all output audio. We did not find significant differences between the baseline and pruned models, except for the UMP and Structured Trimming at 40\% where word-dropping became serious (Figure \ref{fig:syllable}).

\vspace{-10pt}

\vspace{-10pt}

\vspace{-10pt}

\subsection{Model Complexity}
The model sizes and training times are given in Table \ref{tab:complexity}. Differences in training time are minimal except for SNIP, which can be used to accelerate training (up to 3x at 40\% sparsity) by initially removing redundant connections.

\vspace{-5pt}
\begin{table}[h]
    \caption{Model complexity. All models were trained on a single GeForce RTX 3090 GPU.}
    \label{tab:complexity}
    \centering
        \begin{tabular}{l|c|c}
                            & \textbf{Training time (hr)} & \textbf{Model params}     \\
        \midrule
        Baseline / UMP      & 55.4 & \multirow{4}{*}{47.0M}    \\
        PARP                & 55.6 &  \\
        SNIP                & \textbf{40.9} (20\%)  \textbf{19.1} (40\%) & \\
        Sparse Momen.      & 55.2 & \\
        \midrule
        \multirow{2}*{Structured Trim.} & \multirow{2}*{55.4} & \textbf{46.1M} (2\%)   \\
                            & &         \textbf{45.1M} (4\%)
    \end{tabular}
\end{table}
\vspace{-10pt}

\section{Conclusion}

We have presented multiple paradigms of attaining sparsity in TTS models, the first such study for TTS. Compared to previous work \cite{state-sparsity} that shows pruning after training outperforms before or during training, we find that all three approaches are not significantly different for TTS; pre-pruning may have an advantage in shortening training time. Moreover, our experiments illustrate the known difficulty of structured trimming. A possible future direction is to investigate what architectures are amenable to pruning, since our absolute results indicate that maintaining performance at large sparsities on TTS are less achievable than on image classification, which pruning algorithms mainly benchmark against.

\bibliographystyle{IEEEtran}

\bibliography{mybib}


\end{document}